\documentclass[aps,prc,floatfix,nofootinbib,showpacs,superscriptaddress,twocolumn]{revtex4-1}
\usepackage{amsmath}
\usepackage{amssymb}
\usepackage{color}
\usepackage{graphicx}

\definecolor{darkgreen}{rgb}{0,0.5,0}
\definecolor{purple}{rgb}{0.5,0,0.5}
\definecolor{nblue}{rgb}{0.0,0.0,0.50}
\definecolor{scarlet}{rgb}{1.0,0.2,0}

\newcommand{\partialslash}{\mbox{$\not \! \partial$}}

\begin{document}

\title{Confinement contains condensates}

\author{Stanley~J.~Brodsky}
\affiliation{SLAC National Accelerator Laboratory,
Stanford University, Stanford, CA 94309}
\affiliation{Centre for Particle Physics Phenomenology: CP$^3$-Origins, University of Southern Denmark, Odense 5230 M, Denmark}

\author{Craig~D.~Roberts}
\affiliation{Physics Division, Argonne National
Laboratory, Argonne, Illinois 60439, USA}
\affiliation{Department of Physics, Illinois Institute of Technology, Chicago, Illinois}

\author{Robert~Shrock}
\affiliation{C.N. Yang Institute for Theoretical Physics,
Stony Brook University, Stony Brook, NY 11794}

\author{Peter~C.~Tandy} \affiliation{Center for Nuclear Research, Department of
Physics, Kent State University, Kent OH 44242, USA}

\begin{abstract}
Dynamical chiral symmetry breaking and its connection to the generation of
hadron masses has historically been viewed as a vacuum phenomenon.  We argue
that confinement makes such a position untenable.  If quark-hadron duality is a
reality in QCD, then condensates, those quantities that have commonly been viewed as
constant empirical mass-scales that fill all spacetime, are instead wholly
contained within hadrons; i.e., they are a property of hadrons themselves and
expressed, e.g., in their Bethe-Salpeter or light-front wave functions.  We
explain that this paradigm is consistent with empirical evidence, and
incidentally expose misconceptions in a recent Comment.
%
\end{abstract}


\maketitle

\date{9~February 2012}

\section{Background}
\label{sec:Introduction}
Quantum chromodynamics (QCD), as we currently understand it at zero temperature
and zero chemical potential, provides an explanation of the physical, experimentally
observable strong-interaction states; namely, the color-singlet hadrons.  In
the phenomenology of QCD it is conventional to view the quark condensate
$\langle \bar q q \rangle$ as a spacetime-independent constant that fills empty
space.
However, it has long been recognized that in Dirac's light-front (LF) form of relativistic dynamics, which has successfully been applied to QCD \cite{Lepage:1980fj,Brodsky:1997de,Brodsky:2008pg}, the ground state of the theory is a structureless Fock-space vacuum without a $\langle \bar q q \rangle$ condensate, or anything else of this nature.  Furthermore, as was first argued using the LF framework in Ref.\,\cite{Casher:1974xd}, dynamical chiral symmetry breaking (DCSB) and the associated quark condensate must be a property of hadron wave functions, not of the vacuum.  This thesis has also been explored in Refs.\,\cite{Burkardt:1998dd,Brodsky:2008be,Brodsky:2009zd,%
Brodsky:2010xf,Chang:2011mu,Glazek:2011vg}.  One subtlety in characterizing the formal quantity $\langle 0 | {\cal O} | 0 \rangle$, where ${\cal O}$ is a product of quantum field operators, is evident when one recalls that this can automatically be rendered zero by normal-ordering ${\cal O}$.  As we elucidate below, this subtlety is especially delicate in a confining theory because the vacuum state in such a theory is not defined relative to the fields in the Lagrangian, quarks and gluons, but to the actual physical, color-singlet,
states.

In a rigorous statistical mechanical treatment of a phase transition such as that involving magnetism or superconductivity, the transition occurs only in the infinite-volume limit, and the order parameter, e.g., magnetization or Cooper pair condensate, is a constant that extends throughout spacetime. However, as emphasized in Ref.\,\cite{Brodsky:2008be,Brodsky:2009zd}, experimentally one always observes magnetism and superconductivity in finite samples, and the magnetization or Cooper pair condensates are constants only within the material that supports them, not throughout an infinite volume.  In a similar manner, particularly because of confinement, one may argue that QCD condensates are completely contained within that domain which permits the propagation of the gluons and quarks that produce them; namely, inside hadrons.  The conventional view of QCD condensates also has the problem that it predicts a contribution to the cosmological constant that is $10^{46}$ times too large -- a problem that is removed with in-hadron condensates \cite{Brodsky:2009zd,Brodsky:2010xf,Chang:2011mu}.

\section{Condensates and Confinement}
It is worth reiterating that nonzero vacuum expectation values of local
operators in QCD -- the so-called vacuum condensates -- are phenomenological
parameters, which were introduced at a time of limited computational resources
in order to assist with the theoretical estimation of essentially
nonperturbative strong-interaction matrix elements \cite{Shifman:1978bx}.  A
universality of these condensates was assumed; namely, that the properties of
all hadrons could be expanded in terms of the same condensates.  Whilst this
helps to retard proliferation, there are nevertheless infinitely many of them.
As qualities associated with an unmeasurable state (the vacuum) such
condensates do not admit direct measurement.  Practitioners have attempted to
assign values to them via an internally consistent treatment of many separate
empirical observables.  However, only one, the quark condensate, is attributed
a value with any confidence.  The difficulties and capacities of the sum rules
approach are detailed in Ref.\,\cite{Leinweber:1995fn}.

In tackling a problem as difficult as determining the truly observable
predictions of nonperturbative QCD, theory will naturally employ artifices.
Problems arise only when notional elements in the computational structure are
erroneously imbued with an empirical nature.  As noted by a number of authors
\cite{Casher:1974xd,Burkardt:1998dd,Brodsky:2008be,Brodsky:2009zd,Brodsky:2010xf,Chang:2011mu,Glazek:2011vg},
this is the case with the QCD vacuum condensates: from being merely
mass-dimensioned parameters in a theoretical truncation scheme, with no
existence independent of hadrons, in the minds of many they have been
transformed into measurable spacetime-independent vacuum configurations of
QCD's elementary degrees-of-freedom.  In the presence of confinement, the
latter is impossible, and the measurable impact of the so-called condensates is
expressed entirely in the properties of QCD's asymptotically realizable states;
namely hadrons.  Faith in empirical vacuum condensates may be compared with an
earlier misguided conviction that the universe was filled with a luminiferous
aether, which was not overturned before completion of a renowned experiment
\protect\cite{MichelsonMorley}.

It is important to emphasize that confinement is a statement about real-world QCD, in which light-quarks are ubiquitous and pions are light.
Although studies of pure gluonic $SU(3)$ gauge theory, especially via lattice simulations, have given valuable results, such as the spectrum of glueballs (see, e.g., Ref.\,\cite{Morningstar:1999rf} or a recent review \cite{Crede:2008vw}), one must take into account mixing with quarks when relating these to actual QCD.  Similarly, although results obtained with static quark sources, such as Wilson's analytic proof of confinement for strong bare coupling in a lattice gauge theory \cite{Wilson:1974sk}, have been quite valuable, one again must bear in mind that the presence of light quarks in real QCD leads to string-breaking with corresponding meson production.
Confinement is equivalent to exact quark-hadron duality; i.e., that all observable consequences of QCD can be computed using a hadronic basis.  Equivalently, the Hilbert space associated with the measurable Hamiltonian of QCD is spanned by color-singlet state-vectors; viz.,
\begin{equation}
{\cal H}_{\rm QCD} = \sum_n \, E_n |H^{1_c}_n \rangle \langle H^{1_c}_n|\,,
\end{equation}
where $|H^{1_c}_n \rangle$ are color singlets.  Causality entails that QCD
possesses a state of lowest observable energy, which one can choose to be
$E_0=0$.  The state associated with this energy is the vacuum.  It is the state
with zero hadrons.

A precise definition of the vacuum is only possible if one has a nonperturbative definition of the field variable associated with the asymptotic one-particle state, for then the vacuum is that state obtained when the field annihilation operator acts on the asymptotic one-particle state, which is unambiguous.  This is closely connected with the point about normal-ordering.  One may visualize the creation and annihilation operators for such states as rigorously defined via smeared sources on a spacetime lattice.  The ground-state is defined with reference to such operators, employing, e.g., the Gell-Mann--Low theorem \cite{GellMann:1951rw}, which is applicable in this case because there are well-defined asymptotic states and associated annihilation and creation operators.

The notion of a structured vacuum in QCD involves an analogy drawn between dynamical chiral symmetry breaking (DCSB) in the strong interaction and the BCS-theory of superconductivity \cite{Nambu:2011zz}.  The BCS approach is a mean-field theory based on a Hamiltonian expressed in terms of well-defined quasiparticle operators.  There is a known relation between the bare-particle and quasiparticle operators and, under certain conditions, the latter can possess a nonzero expectation value in the vacuum defined via the bare-particle
annihilation operator.  Owing to confinement, these steps are impossible in QCD.
Furthermore, the BCS-based analysis is subject to the comment reiterated above \cite{Brodsky:2008be,Brodsky:2009zd}; namely, that although formally a phase transition in statistical mechanics requires an infinite-volume limit and the resulting order parameter (here the Cooper pair condensate) is a constant throughout infinite space, one actually experimentally observes the Cooper pair condensate to exist only inside finite pieces of superconducting materials, not to be a constant extending throughout infinite space.  In statistical mechanics and condensed matter discussions of real phase transitions and critical phenomena, one is careful to distinguish between the idealized infinite-volume limit and actual experimental observations on finite samples.  One must be equally careful in gauge theories.

Amongst the consequences of confinement is the absence of asymptotic gluon and quark states.  It is therefore impossible to write a valid nonperturbative definition of a single gluon or quark annihilation operator.  To do so would be to answer the question: What is the operator that annihilates a state which is unmeasurable?  So although one can define a perturbative (bare) vacuum for QCD, it is impossible to rigorously define a ground state for QCD upon a foundation of gluon and quark (quasiparticle) operators.  Likewise, it is impossible to construct an interacting vacuum -- a BCS-like trial state -- and hence DCSB in QCD cannot rigorously be expressed via a spacetime-independent coherent state built upon the ground state of perturbative QCD.  Whilst this does not prevent one from following this path to build approximate  models for use in hadron physics phenomenology (Ref.\,\cite{Finger:1981gm} is a pertinent example), it does invalidate any claim that theoretical artifices in such models are accurate descriptions of real QCD.

\section{GMOR Relation}
These remarks provide additional context for the arguments detailed in
Refs.\,\cite{Brodsky:2009zd,Brodsky:2010xf,Chang:2011mu}, which explain that
condensates are localized within hadrons.  Notably, those discussions proceed
via the proof of exact and hence model-independent results in QCD, amongst
them:
the chiral-limit vacuum quark condensate is equivalent to the pseudoscalar
meson leptonic decay constant, in the sense that they are both obtained as the
chiral-limit value of well-defined gauge-invariant hadron-to-vacuum transition
amplitudes that possess a spectral representation in terms of the current-quark
mass \cite{Brodsky:2010xf};
the same is true in the scalar channel \cite{Chang:2011mu};
and in-hadron quark condensates can be represented through a given hadron's scalar form factor at zero momentum transfer \cite{Chang:2011mu}.

It is appropriate here to exemplify these notions via an expression for the in-pseudoscalar-meson quark condensate \cite{Maris:1997hd,Maris:1997tm}:
\begin{equation}
\label{kappazeta}
\kappa^\zeta_{P_{f_1 f_2}} := \rho_{P_{f_1 f_2}}^\zeta f_{P_{f_1 f_2}}\,,
\end{equation}
where $P_{f_1 f_2}$ denotes a pseudoscalar meson comprised of a valence-quark $f_1$ and -antiquark $f_2$, and ($k_\pm = k\pm P/2$)
\begin{eqnarray}
\nonumber
\lefteqn{
i f_{P_{f_1 f_2}} K_\mu = \langle 0 | \bar q_{f_2} \gamma_5 \gamma_\mu q_{f_1} |P \rangle }\\
%
&=&  Z_2\; {\rm tr}_{\rm CD}
\int_{dk}^\Lambda \! i\gamma_5\gamma_\mu S_{f_1}(k_+) \Gamma_{P_{f_1 f_2}}(k;K) S_{f_2}(k_-)\,,  \label{fpigen} \\
%
\nonumber
\lefteqn{ i\rho_{P_{f_1 f_2}}^\zeta = -\langle 0 | \bar f_2 i\gamma_5 f_1 |P \rangle}\\
&=& Z_4\; {\rm tr}_{\rm CD}
\int_{dk}^\Lambda \! \gamma_5 S_{f_1}(k_+) \Gamma_{P_{f_1 f_2}}(k;K) S_{f_2}(k_-) \,.\label{rhogen}
\end{eqnarray}
Here $\int_{dk}^\Lambda$ is a Poincar\'e-invariant regularization of the
integral, with $\Lambda$ the ultraviolet regularization mass-scale,
$Z_{2,4}(\zeta,\Lambda)$ are renormalization constants, with $\zeta$ the
renormalization point, $\Gamma_{P_{f_1 f_2}}$ is the meson's Bethe-Salpeter
amplitude, and $S_{f_1,f_2}$ are the component dressed-quark propagators, with
$m_{f_1,f_2}$ the associated current-quark masses.  Equation~(\ref{fpigen})
describes the pseudoscalar meson's leptonic decay constant; i.e., the
pseudovector projection of the meson's Bethe-Salpeter wave-function onto the
origin in configuration space; Eq.\,(\ref{rhogen}) describes its pseudoscalar
analogue; and $m_{P_{f_1 f_2}}$ is the meson's mass.

It is an exact result in QCD, valid for arbitrarily small or large current-quark masses and for both ground- and excited-states \cite{Holl:2004fr}, that
\begin{equation}
\label{GMORP}
f_{P_{f_1 f_2}}^2 m_{P_{f_1 f_2}}^2 = [m_{f_1}^\zeta +m_{f_2}^\zeta]\, \kappa^\zeta_{P_{f_1 f_2}}
= [\hat m_{f_1} + \hat m_{f_2}]\, \hat \kappa_{P_{f_1 f_2}},
%
%
\end{equation}
where the circumflex indicates a renormalization-group-invariant quantity.  Moreover \cite{Maris:1997hd}
\begin{equation}
\lim_{\hat m\to 0} \kappa^\zeta_{P_{f_1 f_2}}
=
Z_4 \, {\rm tr}_{\rm CD}\int_{dq}^\Lambda S^0(q;\zeta) =  -\langle \bar q q \rangle_\zeta^0\,;
\label{qbqpiqbq0}
\end{equation}
namely, that the so-called vacuum quark condensate is, in fact, the
chiral-limit value of the in-meson condensate; i.e., it describes a property of
the chiral-limit pseudoscalar meson.  This condensate is therefore no more a
property of the ``vacuum'' than the pseudoscalar meson's chiral-limit leptonic
decay constant.  Moreover, given that Eq.\,\eqref{qbqpiqbq0} is an identity in
QCD, any veracious calculation of $\langle \bar q q \rangle_\zeta^0$ is the
computation of a gauge-invariant property of the pion's wave-function.

It is also valuable to highlight the precise form of the Gell-Mann--Oakes--Renner (GMOR) relation; viz., Eq.\,(3.4) in Ref.\,\cite{GellMann:1968rz}:
\begin{equation}
\label{gmor}
m_\pi^2 = \lim_{P^\prime \to P \to 0} \langle \pi(P^\prime) | {\cal H}_{\chi{\rm sb}}|\pi(P)\rangle\,,
\end{equation}
where $m_\pi$ is the pion's mass and ${\cal H}_{\chi{\rm sb}}$ is that part of the hadronic Hamiltonian density which explicitly breaks chiral symmetry.  It is crucial to observe that the operator expectation value in Eq.\,(\ref{gmor}) is evaluated between pion states.
Moreover, the virtual low-energy limit expressed in Eq.\,\eqref{gmor} is purely formal.  It does not describe an achievable empirical situation, as we explain in connection with Eq.\,\eqref{eq:constituents} below.

In terms of QCD quantities, Eq.\,(\ref{gmor}) entails
\begin{eqnarray}
\label{gmor1}
\lefteqn{
\forall m_{ud} \sim 0\,,\;  m_{\pi^\pm}^2 =  m_{ud}^\zeta \, {\cal S}_\pi^\zeta(0)\,,}\rule{7.2em}{0ex}\\
{\cal S}_\pi^\zeta(0) & = & - \langle \pi(P) | \mbox{\small $\frac{1}{2}$}(\bar u u + \bar d d) |\pi(P)\rangle\,,
\label{gmor1a}
\end{eqnarray}
where $m_{ud}^\zeta = m_u^\zeta+m_d^\zeta$, $m_{u,d}^\zeta$ are the
current-quark masses, and ${\cal S}^\zeta(0)$ is the pion's scalar form factor
at zero momentum transfer, $Q^2=0$.  The right-hand-side (rhs) of
Eq.\,(\ref{gmor1}) is proportional to the pion $\sigma$-term (see, e.g.,
Ref.\,\cite{Flambaum:2005kc}).  Consequently, using the connection between the
$\sigma$-term and the Feynman-Hellmann theorem, Eq.\,(\ref{gmor}) is actually
the statement
\begin{equation}
\label{pionmass2}
\forall m_{ud} \sim 0\,,\; m_\pi^2 = m_{ud}^\zeta \frac{\partial }{\partial m^\zeta_{ud}} m_\pi^2.
\end{equation}

Now, using Eq.\,\eqref{GMORP}, one obtains
\begin{equation}
\label{gmor2}
{\cal S}_\pi^\zeta(0)
= \frac{\partial }{\partial m^\zeta_{ud}} m_\pi^2
=\frac{\partial }{\partial m^\zeta_{ud}} \left[ m_{ud}^\zeta\frac{\rho_\pi^\zeta}{f_\pi}\right].
\end{equation}
Equation~(\ref{gmor2}) is valid for any values of $m_{u,d}$, including the neighborhood of the chiral limit, wherein
\begin{equation}
\label{gmor3}
\frac{\partial }{\partial m^\zeta_{ud}} \left[ m_{ud}^\zeta\frac{\rho_\pi^\zeta}{f_\pi} \right]_{m_{ud} = 0}
= \frac{\rho_\pi^{\zeta 0}}{f_\pi^0}\,.
\end{equation}
The superscript ``0'' indicates that the quantity is computed in the chiral limit.  With Eqs.\,\eqref{kappazeta}, \eqref{qbqpiqbq0}, (\ref{gmor1}), (\ref{gmor2}), (\ref{gmor3}), one has shown that in the neighborhood of the chiral limit
\begin{equation}
m_{\pi^\pm}^2 =  -m_{ud}^\zeta  \frac{\langle \bar q q \rangle^{\zeta 0}}{(f_\pi^0)^2} + {\rm O}(m_{ud}^2).
\end{equation}
This is a QCD derivation of the commonly recognized form of the GMOR relation.  Neither PCAC nor soft-pion theorems were employed in analyzing the rhs of Eqs.\,\,\eqref{gmor}, (\ref{gmor1}).

This recapitulation of the analysis in Ref.\,\cite{Chang:2011mu} emphasizes
anew that any connection between the pion mass and a vacuum quark condensate is
purely a theoretical artifice.  The true connection is that which one would
expect; viz., the pion's mass is a property of the pion, determined by the
interactions between its constituents.

\section{Consistency with Empirical Information}
\label{sec:empirical}
Notwithstanding the strength of the arguments in
Refs.\,\cite{Brodsky:2009zd,Brodsky:2010xf,Chang:2011mu}, and those we have
expressed above in addition, the conventional view of the chiral order
parameter is so well entrenched that they have been questioned
\cite{Reinhardt:2012xs}.  This presents us with a welcome opportunity to
articulate additional features of our perspective, in addition to dispelling
the misapprehensions expressed therein.

Plainly, the notion of in-hadron condensates does not contradict any empirical observation, for it may be embedded in a broader context by considering just what is observable in quantum field theory \cite{Weinberg:1978kz}: ``\ldots although individual quantum field theories have of course a good deal of content, quantum field theory itself has no content beyond analyticity, unitarity, cluster decomposition and symmetry.''  Our arguments exploit these facts.
If QCD is a confining theory, then the principle of cluster decomposition is only realized for color singlet states \cite{Krein:1990sf} and thus the only vacuum upon which gluon or quark field operators can be defined to act is that within the hadron they constitute.

\begin{figure}[t]

\includegraphics[clip,width=0.4\textwidth]{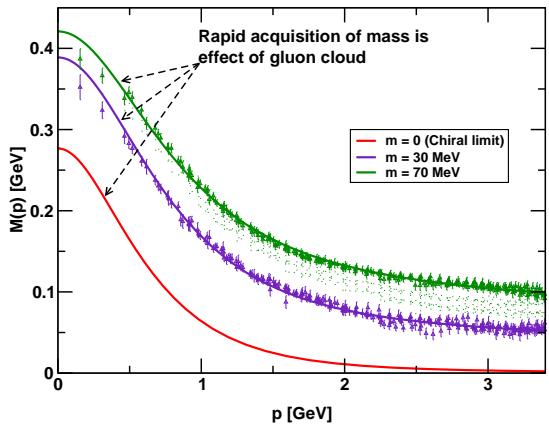}
\caption{\label{gluoncloud} Dressed-quark mass function, $M(p)$: \emph{solid
curves} -- Dyson-Schwinger equation (DSE) results,
\protect\cite{Bhagwat:2003vw,Bhagwat:2006tu,Bhagwat:2007vx}, ``data'' -- lattice-QCD
simulations \protect\cite{Bowman:2005vx}.  (N.B.\ $m=70\,$MeV is the uppermost
curve.  Current-quark mass decreases from top to bottom.)  
bulk of the light-quark constituent mass arises from gluons through DCSB.  The
constituent mass arises from a cloud of low-momentum gluons attaching
themselves to the current-quark: DCSB is a truly nonperturbative effect that
generates a running mass for quarks even when the
renormalization-point-invariant current-quark mass vanishes, as evidenced by
the $m=0$ curve.}
\end{figure}


It is manifestly mistaken to suggest \cite{Reinhardt:2012xs} that containing condensates within hadrons entails that the lowest excitations in the pseudoscalar and scalar channels are degenerate.  In QCD, as illustrated in Fig.\,\ref{gluoncloud}, DCSB is expressed in the dressed-quark mass-function through the material enhancement of $M(p)$ on the domain of infrared momenta, $p\lesssim 5\Lambda_{\rm QCD}$.  In modern approaches to the bound-state problem, such as DSE- and lattice-QCD, bound-states are constituted using such propagators. Bound-states therefore express the features realized in these propagators.  It is possible to illustrate this using the simplest of confining models.  With a symmetry-preserving and confining regularization of a vector-vector contact-interaction, the dressed-quark is described by a dynamically generated mass, $M$, which is large in the chiral limit.
(For some early work showing dynamical fermion mass generation via analysis of DSEs at
sufficiently strong coupling, see Refs.\,\cite{Lane:1974he,Politzer:1976tv,Holdom:1984sk,Yamawaki:1985zg,Appelquist:1986an}.  A general argument that confinement in a QCD-like theory implies the generation of a dynamical quark mass was given in \cite{Casher:1979vw}.  Early discussions of the relation between the dynamically generated fermion mass and $f_\pi$ include Refs.\,\cite{Jackiw:1973tr,Pagels:1979hd}.)
At lowest-order in a symmetry-preserving truncation of the model's DSEs \cite{Bender:1996bb}, one finds algebraically that $m_\pi= 0$, $m_\sigma = 2 M$ \cite{Roberts:2011wy}.  Corrections to the leading order truncation do not change $m_\pi$ but markedly increase $m_\sigma$ \cite{Chang:2009zb}.  Degeneracy of the lowest excitations in the pseudoscalar and scalar channels is only achieved when $M=0$; i.e., if chiral symmetry is not dynamically broken.  The splitting between vector and axial-vector mesons and parity partners in the baryon spectrum are explained in the same manner \cite{Chang:2011ei,Roberts:2011cf}.

The spectrum of QCD exhibits a large splitting between parity partners because
of DCSB, which is manifest in the Schwinger functions of the excitations
confined within hadrons.  It should be recognized that the computation of
color-nonsinglet Schwinger functions in isolation is an artifice.  In the fully
self-consistent treatment of bound-states, the dressing phenomenon illustrated
in Fig.\,\ref{gluoncloud} takes place in the background field generated by the
other constituents.  Its influence is concentrated in the far infrared,
$p\lesssim \Lambda_{\rm QCD}$, and its presence ensures the manifestations of
gluon- and quark-dressing are gauge invariant.

Apropos this discussion, we recapitulate here some remarkable features of the
pseudoscalar meson bound-state problem.  Since the dressed-quark propagator has
a spectral representation when considered as a function of current-quark mass
\cite{Langfeld:2003ye}, one can derive from the axial-vector Ward-Takahashi
identity a collection of Goldberger-Treiman-like relations for the pion
\cite{Goldberger:1958vp,Maris:1997hd,Bhagwat:2007ha}.  Of particular relevance
herein is \begin{equation}
\label{bwti}
E_\pi(k;0) = \frac{1}{f_\pi^0} \,B_0(k^2)\,,
\end{equation}
where $E_\pi$ is that Lorentz-invariant function in the pion's Bethe-Salpeter
amplitude which multiplies $\gamma_5$, $B_0$ is the scalar part of the
dressed-quark self-energy computed with vanishing current-quark mass, and
$f_\pi^0$ is the pion's leptonic decay constant in that limit.
Equation~\eqref{bwti} can be used to prove that a massless pseudoscalar meson
appears in the chiral-limit spectrum if, and only if, chiral symmetry is
dynamically broken \cite{Maris:1997hd}.  Moreover, it exposes the fascinating
consequence that the solution of the two-body pseudoscalar bound-state problem
is almost completely known once the one-body problem is solved for the
dressed-quark propagator, with the relative momentum within the bound-state
identified unambiguously with the momentum of the dressed-quark.  This latter
emphasizes that Goldstone's theorem has a pointwise expression in QCD, and that
expression is contained within hadrons because dressed-quarks are confined.

This brings us to another point that is worth elucidating.  In learning that
the so-called vacuum quark condensate is actually the chiral-limit value of an
in-pion property, some respond as follows.
The electromagnetic radius of any hadron which couples to pseudoscalar mesons
must diverge in the chiral limit.  This long-known effect arises because the
propagation of \emph{massless} on-shell color-singlet pseudoscalar mesons is
undamped \cite{Beg:1973sc,Pervushin:1974nm,Gasser:1983yg,Alkofer:1993gu}.
Therefore, does not each pion grow to fill the universe; so that, in this
limit, the in-pion condensate reproduces the conventional paradigm?

Confinement, again enables one to refute this objection.  As noted above, general arguments, as well as DSE- and lattice-QCD studies, indicate that confinement entails dynamical mass generation for quarks.
In a vectorial gauge theory such as QCD, a quark mass term in the action is automatically gauge invariant.  Arguments have also been given that confinement entails an effective mass for gluons (effective in the sense that it is not associated with a term in the action that violates color gauge invariance).
These observations have often been explained, most recently in Sec.\,III of
Ref.\,\cite{Bashir:2012fs}.
The zero-momentum value of the momentum-dependent dynamical quark masses $M(0)$ and effective gluon mass $m_g(0)$ remain large in the limit of vanishing current-quark mass.  In fact, these values are almost independent of the current-quark mass in the neighborhood of the chiral limit.  (This is apparent in Fig.\,\ref{gluoncloud}.)
As a consequence, one can argue that the quark-gluon containment-radius of all hadrons is finite in the chiral limit.  Indeed it is almost insensitive to the magnitude of the current-quark mass because the dynamical masses of the hadron's constituents are frozen at large values; viz.,
\begin{equation}
\label{eq:constituents}
M(0) \lesssim m_g(0) =: m_{\rm c} \sim 2\Lambda_{\rm QCD}-3 \Lambda_{\rm QCD}
\,.
\end{equation}
These considerations suggest that the divergence of the electromagnetic radius does not correspond to expansion of a condensate from within the pion but rather to the copious production and subsequent propagation of composite pions, each of which contains a condensate whose value is essentially unchanged from its nonzero current-quark mass value within a containment-domain whose size is similarly unaffected.  That domain is specified by a radius
$r_{\rm c} \sim 1/m_{\rm c}$.

There is more to be said in connection with the definition and consequences of a chiral limit.  Nambu-Goldstone bosons are weakly interacting in the infrared limit.  However, at nonzero energies, their interactions are, in general, strong, and they always couple strongly, e.g., to the nucleon.  Plainly, the existence of strongly-interacting massless composites would have an enormous impact on the evolution of the universe; and it is naive to imagine that one can simply set $\hat m_{u,d}=0$ and consider a circumscribed range of manageable consequences whilst ignoring the wider implications for hadrons, the Standard Model and beyond.  For example, with all else held constant, Big Bang Nucleosynthesis is very sensitive to the value of the pion-mass \cite{Flambaum:2007mj,Bedaque:2010hr}.  We are fortunate that the absence of quarks with zero current-quark mass has produced a universe in which we exist so that we may carefully ponder the alternative.

As we mentioned in the Introduction, a universality of condensates was assumed
in order to slow growth in the number of undetermined parameters that appear in
the sum rules scheme.  As grasped in Ref.\,\cite{Reinhardt:2012xs}, with the
appreciation that condensates are contained within hadrons, the assumption of
universality is seen to be quantitatively false \cite{Chang:2011mu}.  This is
similar, in fact, to the assumption of vacuum saturation for the four-quark
condensate, which underestimates the correct result by $\sim65$\%
\cite{Nguyen:2010yj}.  It is nonetheless interesting that the magnitude of the
in-hadron quark condensate is only weakly sensitive to the host state
\cite{Chang:2011mu,Roberts:2011ea}.  This, too, is tied to the preeminent role
played by the dressed-gluon and -quark propagators in producing bound-states
and their masses.

On the other hand, the suggestion \cite{Reinhardt:2012xs} that the containment
of condensates with hadrons precludes chiral symmetry restoration at nonzero
baryon density is spurious.  As reviewed in Ref.\,\cite{Roberts:2000aa}, a
nonzero chemical potential, $\mu$, has a dramatic impact on the dressed-quark
propagator once $\mu$ arrives at the vicinity of a critical value.  At zero
temperature that value is $\mu_{\rm cr} \approx 0.3\,$GeV \cite{Qin:2010nq}.
This is also true of nonzero temperature; and as the realization of DCSB in the
dressed-quark propagator is suppressed and finally eliminated by growth of
these intensive thermodynamic parameters, so do parity partners become
degenerate.  A concrete example in a confining model is detailed in
Ref.\,\cite{Maris:2000ig}.
We note in addition that merely to conjecture \cite{McLerran:2007qj} a phase of matter that exhibits confinement but simultaneously manifests chiral symmetry does not establish its existence; and, in fact, available confining models for QCD's gap equation do not support this speculation \cite{Qin:2010nq}.

\section{Analysis within a Model}
Owing to the importance of DCSB, a complete understanding of hadron structure
is only possible if the properties of meson and baryon ground- and
excited-states can be correlated within a single symmetry-preserving framework,
where symmetry-preserving means that all relevant Ward-Takahashi identities are
satisfied and Poincar\'e covariance is respected.  In addition, the framework
must treat mesons and baryons on an equal footing, and be applicable to all
mesons and baryons.
These conditions are satisfied, in principle, by the light-front approach to
hadron structure and interactions \cite{Brodsky:2004er};
presumably by lattice-QCD, once the various necessary limits are carefully taken;
and plainly by the more intuitive, less computationally intensive DSE approach .

The DSE framework expresses the results of perturbative QCD and provides a unified treatment of, amongst other things:
meson and baryon spectra \cite{Eichmann:2008ef,Chang:2011ei};
hadron electromagnetic elastic and transition form factors \cite{Maris:2000sk,Roberts:2010rn,Eichmann:2011ej};
meson-meson scattering \cite{Bicudo:2001aw,Bicudo:2001jq,Cotanch:2002vj};
and the distribution functions that arise in analyses of deep inelastic scattering \cite{Hecht:2000xa,Holt:2010vj,Aicher:2010cb,Nguyen:2011jy}.
The expression and realization of in-hadron condensates has most widely been
elucidated within this framework precisely because it is applicable to all
hadrons and treats all hadrons equally; viz., as Poincar\'e-covariant
bound-states of confined, dressed-partons.
N.B.\ In the DSE- and lattice-QCD computation of the properties of an isolated hadron, all QCD's dynamical content is expressed within that hadron, owing to confinement.

Historically; i.e., before the nature of DCSB and its intimate connection with gluon and quark dressing was fully understood, it was popular to employ simple models that cannot treat all hadrons equally.
This is the case with the mean-field soliton model used in Ref.\,\cite{Reinhardt:2012xs}.  Amongst its weaknesses, the model is not Poincar\'e covariant and hence not symmetry preserving;
it is not confining, the understanding of which is critical to the realization
of in-hadron condensates;
it does not treat mesons and baryons on an equal footing, with, e.g., pseudoscalar and scalar mesons being pointlike but the nucleon having nonzero extent;
it is not applicable to all mesons and baryons, e.g., requiring material amendment if a description of vector- and higher-mass-mesons is to be attempted;
and it produces valence-quark distribution functions in marked disagreement with QCD \cite{Holt:2010vj}.
In addition to these factors, the analysis in Ref.\,\cite{Reinhardt:2012xs} is flawed.  As we now explain, their extraction of a quark condensate associated with the soliton solution is neither model-consistent nor correct.



Consider the two-flavor model defined via the following generating functional
\begin{equation}
Z[J] = \int D \bar \psi D \psi \,{\rm exp}  \Big\{i\! \int\! d^4x \, \bar \psi (i \partialslash - m_0  - J ) \psi  +
i{\cal A}_{\rm int} \Big\} ,
\label{eq:Z}
\end{equation}
where $\psi$ is a fermion field, $J(x)$ is a scalar-isoscalar source function that monitors the local composite operator $\bar \psi(x) \psi(x) $, and ${\cal A}_{\rm int}$ is a standard $U(2)\times U(2)$-symmetric pseudoscalar and scalar 4-fermion contact interaction.  In this model it is formally true that
\begin{equation}
\langle \bar \psi \psi \rangle = \frac{i\delta}{\delta J(x)} \, {\rm ln} Z[J]  \, |_{J = 0} \,.
\label{eq:qbarq}
\end{equation}
However, this expression is meaningless unless a regularization and renormalization scheme is specified.  Extreme care must in particular be employed when the current-mass is nonzero \cite{Langfeld:2003ye}.  In order to track contributions to $\langle \bar \psi \psi \rangle$ that are associated with the final hadronic application of the model we note that, through all manipulations, the extension \mbox{$i \partialslash_x \to i \partialslash_x -J(x) $} will identify all contributions to the required derivative.

In order to proceed, it is customary \cite{Cahill:1985mh} to eliminate the four-fermion interaction in favor of pseudoscalar and scalar auxiliary fields that are linearly coupled to the fermions, a procedure that in the present case leads to
\begin{eqnarray}
\nonumber
\lefteqn{   {\cal A}_{\rm B}[S,P] =  - i {\rm Tr}_{\bar\Lambda}\,  {\rm ln}\, [ i \partialslash - (S + i\gamma_5 P) ] }\\
&&  - \frac{1}{4G} \int d^4 x \;  {\rm tr}_{\rm F} [(S - m_0)^2 + P^2]  \, ,
 \label{eqn:Action_B}
\end{eqnarray}
where the flavor-singlet component of the field $S$ is defined to absorb the current-quark  mass, $m_0$.  The subscript $\bar\Lambda$ indicates that a regularization scheme has been introduced, characterized by a mass-scale $\bar\Lambda \sim 0.6\,$GeV, which is a parameter in this non-renormalizable model: the details are not important here.  Under $U(2)\times U(2)$ transformations, $S^a$ transforms as $\bar \psi \tau^a \psi$ and $P^a$ as  $\bar \psi \tau^a \gamma_5 \psi$, where $\tau^0 = \mathbf{1}_{2\times 2}$ and $\{\vec\tau\}$ are the Pauli matrices.

At this point a stationary phase approximation is explored in connection with $Z[J]$ and the extremum of ${\cal A}_{\rm B}$ is argued to produce the model's vacuum configuration.  (This is equivalent to the mean-field approximation in Ref.\,\cite{Finger:1981gm}.)  It is expressed through the rainbow-ladder truncation of the model's gap equation, and produces $P_0=0$ and $S_0= m$, the dressed-fermion mass.  The properties and interactions of meson-like fluctuations about the extremum are described by
\begin{equation}
\tilde {\cal A}_{\rm B}[\tilde S, \tilde P] =  {\cal A}_{\rm B}[S,P] - {\cal A}_{\rm B}[S_0,0] \, ,
\label{eq:mesonaction}
\end{equation}
where $\tilde S = S - S_0$, $\tilde P = P - P_0$.
At second-order of an expansion in terms of field derivatives, $\tilde {\cal A}_{\rm B}$ reproduces a linear $\sigma$-model with an effective potential in the form of the so-called Mexican hat.  (The precise expression and meaning of the Mexican hat potential in QCD are explained in Ref.\,\cite{Chang:2009at}.)  The model displays DCSB but not confinement, and the auxiliary fields represent pointlike mesons described by the rainbow-ladder Bethe-Salpeter equation. (Confinement may be implemented \cite{Frank:1991qp,Frank:1992xi} via modifications that express the quark dressing illustrated in Fig.\,\ref{gluoncloud}.)

We note that Eq.\,\eqref{eq:mesonaction} actually contains a fermion determinant.  It is therefore nonlocal, possessing meson-field derivatives of arbitrarily high order; i.e., infinitely many meson self-couplings in excess of the quadratic, cubic and quartic terms that are the only ones usually retained.
Moreover, the linear sigma model, and its generalization here, produce a spacetime-independent extremum for the scalar field, $S_0=m$, and through this device describe DCSB and meson-fermion physics.  This spacetime-independent field is also the mass of the model's dressed-fermion.  It is a constant, in striking contrast to the dressed-quark mass-function in QCD, Fig.\,\ref{gluoncloud}.  Thus, from the outset the model exhibits a marked and empirically verifiable departure from QCD; and hence the associated unbounded three-space support for the energy-density associated with the extremum should not be misconstrued as a feature of QCD.

The fermion condensate associated with the model's mean-field ground-state is
\begin{equation}
\langle \bar \psi \psi \rangle_{\rm sp} =  \frac{i\delta}{\delta J(x)}  Tr_{\bar\Lambda} {\rm ln}
[ i \partialslash - m  - J ] \Big |_{J=0} \, ,
\label{eq:saddle-pt-cond}
\end{equation}
and this yields the following expression in terms of the Euclidean-space fermion propagator
\begin{equation}
\langle \bar \psi \psi \rangle_{\rm sp} = -  {\rm tr}_{\rm CD}\int^{\bar\Lambda} \!\!\!\! \mbox{\footnotesize $\displaystyle\frac{d^4 q}{(2\pi)^4}$} \,G(q) \, .
\label{eq:trace_S}
\end{equation}
Here the regularization matters.  As implemented in
Ref.\,\cite{Reinhardt:2012xs}, it is simply a proper-time scheme, which is a
typical procedure within the model's milieu.  However, it yields a result that
has no connection with the three common, valid expressions of the quark
condensate in QCD, all of which are precisely equivalent to the chiral-limit
value of the in-pseudoscalar-meson condensate, Eq.\,\eqref{kappazeta}
\cite{Brodsky:2010xf,Chang:2011mu,Langfeld:2003ye}.  Indeed, first pointed out
in Refs.\,\cite{Lane:1974he,Politzer:1976tv}, the quantity commonly called the
vacuum quark condensate may be read from the coefficient of the $1/q^4$ term in
the operator product expansion of the chiral-limit dressed-quark propagator.
Even within the soliton model, the left-hand-side of Eq.\,\eqref{eq:trace_S}
has no connection with that quantity.  Equation~\eqref{eq:trace_S} is therefore
a model-specific definition that violates known relations in QCD.  On the other
hand, Eq.\,\eqref{qbqpiqbq0} is precise in QCD.

There has hitherto been no explicit mention of a baryon.  At this point Ref.\,\cite{Reinhardt:2012xs} proceeds by developing a mean-field approximation to the expectation value of a three fermion propagator based on the action just described.  The general procedure was explained in Ref.\,\cite{Williams:1983bf}.  The result is a non-topological soliton model described by the action
\begin{equation}
\Gamma_{\rm sol}[\tilde S, \tilde P] = {\cal A}_{\rm val}[\tilde S, \tilde P] + \tilde {\cal A}_{\rm B}[\tilde S, \tilde P] \, ;
\label{eq:Gamma}
\end{equation}
where the precise form of the valence quark piece, ${\cal A}_{\rm val}$, is not relevant here.  The mean field equations of motion are
\begin{equation}
\frac{\delta \Gamma_{\rm sol} }{\delta \tilde S, \tilde P} = 0\,.
\end{equation}

The mean-field method seeks a time-independent solution with time-independent fields:  \begin{equation}
\Gamma_{\rm sol} = (-\int dt) \int d^3x \,\left[ {\cal E}_{\rm sol}(\vec{x}) =
{\cal E}_{\rm val}(\vec{x})+{\cal E}_{\rm B}(\vec{x})\right],
\end{equation}
with derived energy densities ${\cal E}_{\rm val}(\vec{x})$, ${\cal E}_{\rm B}(\vec{x})$ being of finite three-space extent.  This is satisfied by composite three-valence-fermion states bound with respect to their constituents' masses; viz., $m_{3\psi}<3m$.  Such states act as a finite-size source that induces non-zero values for $\tilde S(\vec{x}) = S - S_0$ and $\tilde P(\vec{x}) = P - P_0$ over a commensurate three-space volume.
N.B.\ In order to achieve the physical result that the soliton energy-density produced by the fields $\tilde S, \tilde P$ does not exist outside of the soliton, it is critical to subtract the extremum-value of the action when defining \mbox{$\tilde {\cal A}_{\rm B}[\tilde S, \tilde P]$}.

Through the simple expedient of the replacement \mbox{$i \partialslash \to i \partialslash -J $}, the expression for $\langle \bar \psi \psi \rangle$ in Eq.\,\eqref{eq:qbarq} may be traced through the mean-field development to find
\begin{equation}
\langle \bar \psi \psi \rangle_{\rm sol} = - \frac{\delta \Gamma_{\rm sol} }{\delta J(x)} \Big |_{J=0} \, .
\label{eq:cond_sol}
\end{equation}
Since the three-space density of the model's action vanishes outside the soliton, it is impossible for an internally consistent calculation of $\langle \bar \psi \psi \rangle_{\rm sol}$ to produce a result that has support in a three-space volume that is significantly larger than that of the action/energy density.

Since ${\cal A}_{\rm val}$ is built on the fermion kinetic term and $\tilde {\cal A}_{\rm B}$ contains a subtracted logarithm of the fermion determinant, both terms in $\Gamma_{\rm sol}$  will contribute to $\langle \bar \psi \psi \rangle_{\rm sol}$.
However, outside the soliton, which is most relevant herein, there is only $\tilde {\cal A}_{\rm B}$, so let's focus on its contribution to Eq.\,\eqref{eq:qbarq}.  In the large-distance limit the internally consistent model result is
\begin{equation}
\lim_{|\vec{x}| \to \infty} \mbox{\footnotesize $\displaystyle\frac{i\delta}{\delta J(x)}  {\rm Tr}_{\bar\Lambda} $}\Big\{ {\rm ln}
[ i \partialslash - S - i\gamma_5 P  - J ] -   {\rm ln} [ i \partialslash - m  - J ]   \Big\} = 0 \, .
\label{eq:cond_B}
\end{equation}
Before an extremum was defined and subtracted, $\delta/\delta S(x)$ could be substituted for $\delta/\delta J(x)$ in treating Eq.\,\eqref{eq:qbarq} so long as one only applied $\delta/\delta S(x)$ to the first term on the right-hand-side of Eq.\,\eqref{eqn:Action_B}.  However, this shift is invalid if one first expresses the extremum at \mbox{$S_0 = m$} and treats the subtraction term as a constant that does not respond to variations in the source associated with $\langle\bar\psi\psi\rangle$.  That procedure leads to the following erroneous result
\begin{equation}
\lim_{|\vec{x}| \to \infty} \mbox{\footnotesize $\displaystyle\frac{i\delta}{\delta S(x)}  Tr_{\bar\Lambda}$} \Big\{ {\rm ln}
[ i \partialslash - S - i\gamma_5 P] -  {\rm ln} [ i \partialslash - m ] \Big\} =  \langle \bar \psi \psi \rangle_{\rm sp} \, ;
\end{equation}
and this is the mathematical mistake made in Ref.\,\cite{Reinhardt:2012xs}.

With use of the consistent procedure involving $\delta/\delta J(x)$ throughout, one obtains a condensate with bounded support, localized within a finite volume of three-space, whose profile tracks that of the soliton's energy/action density.   Following this procedure, the curves attributed to the soliton's fermion condensate in Figs.\,1,2 of Ref.\,\cite{Reinhardt:2012xs} are shifted upwards by an amount that produces zero outside the soliton.

\section{View from the Light-Front}

The subtraction identified here via the calculus of functional integrals may also be seen to arise naturally through normal-ordering in the equivalent procedure of second quantization.  For the energy of the soliton state one would deal with \mbox{$\langle {\cal S} |  : \!{\cal H}_{\cal S} \!: | {\cal S} \rangle $}, where ${\cal H}_{\cal S}$ is the model Hamiltonian and ${\cal S}$ denotes the soliton state.   The importance of normal-ordering in discussing the connection between physical and vacuum matrix elements, especially for condensates, is emphasized in Ref.\,\cite{Brodsky:2009zd}

We note that, in general, normal-ordering in the equal-time second-quantized formulation of  a quantum field theory which exhibits essentially nonperturbative phenomena is an ill-defined operation because, e.g., the exponentiation involved in writing the Heisenberg field operator induces all orders of bare parton creation and annihilation processes, and no finite sum can recover a nonperturbative effect.  As indicated in the Introduction, a mean-field approximation to a non-confining theory can be used to define a nonperturbative but truly approximate set of states that provides a diagonal basis and associated single quasiparticle creation and annihilation operators.  However, this is impossible for the confined gluons and quarks of QCD.

Of course, the question of normal-ordering is eliminated if one employs the light-front formulation of quantum field theory.  In that case the vacuum is defined as the lowest-mass eigenstate of the associated light-front Hamiltonian by quantizing at fixed $\tau= t-z$; and this vacuum is remarkably simple because the kinematic restriction to $k^+>0$ ensures that the ground-state of the interacting Hamiltonian is the same as that of the free Hamiltonian.
There are other advantages, too.  The front-form vacuum and its eigenstates are Lorentz invariant, whereas the instant form vacuum depends on the observer's Lorentz frame.  And the instant form vacuum is a state defined at the same time, $t$, at all spatial points in the universe, whereas the front-form vacuum senses only those phenomena which are causally connected; i.e., within an observer's light-cone.

This last point ensures that the front-form is well-suited to computation of
the cosmological constant because the constant is a property of the universe
measured within the causal horizon; i.e., it is expressed in the matrix element
of the energy-momentum tensor in the background universe, which is completely
determined by events that occur within a causally connected domain.  It is
practically impossible, on the other hand, to obtain a reliable result using
instant-form dynamics since the truncations necessary in order to obtain a
result will generally violate Lorentz invariance.  Hence one should not be
surprised when expectations based on assumed properties of the vacuum
associated with a truncated instant-form Hamiltonian are misleading.

\begin{figure}[t]
\includegraphics[clip,width=0.32\textwidth]{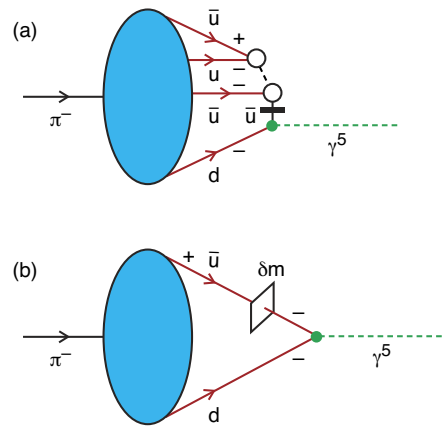}
\caption{\label{instantaneous}
Light-front contributions to $\rho_\pi=-\langle 0| \bar q \gamma_5 q |\pi\rangle$, $f_\pi \rho_\pi$ is the in-pion condensate.
\emph{Upper panel} -- A non-valence piece of the meson's light-front wave-function, whose contribution to $\rho_\pi$ is mediated by the light-front instantaneous quark propagator (vertical crossed-line).  The ``$\pm$'' denote parton helicity.
\emph{Lower panel} -- There are infinitely many such diagrams, which can introduce chiral symmetry breaking in the light-front wave-function in the absence of a current-quark mass.
(The case of $f_\pi$, which is also an order parameter for DCSB, is analogous.)}
\end{figure}

With the shift to a paradigm in which DCSB is expressed as an in-hadron property, one can readily visualize a mechanism that might produce DCSB within the light-front formulation of QCD.  As illustrated in Fig.\,\ref{instantaneous}, the light-front-instantaneous quark propagator can mediate a contribution from higher Fock state components to the matrix elements
\begin{eqnarray}
\nonumber
f_\pi P^- & = & 2 \sqrt N_c \, Z_2 \int_0^1\! dx \! \int\mbox{\footnotesize $\displaystyle\frac{d^2 k_\perp}{16\pi^3}$} \, \psi(x,k_\perp) \frac{k_\perp^2\! + m_\zeta^2}{P^+\,x(1-x)}\\
&& + \mbox{instantaneous}\,,
\label{LFfpi} \\
\nonumber
\rho_\pi & = & \sqrt N_c \, Z_2 \int_0^1\! dx\!
\int\mbox{\footnotesize $\displaystyle\frac{d^2 k_\perp}{16\pi^3}$} \,
\psi(x,k_\perp) \, \frac{ m_\zeta }{x (1-x)}\\
&& + \mbox{instantaneous}\,,
\label{LFrhopi}
\end{eqnarray}
where $P=(P^+,P^-=m_\pi^2/P^+,\vec{0}_\perp)$ and both currents receive contributions from the ``instantaneous'' part of the quark propagator ($\sim \gamma^+/k^+$) and the associated gluon emission, which are not written explicitly.  In Eqs.\,(\ref{LFfpi}), (\ref{LFrhopi}), $\psi(x,k_\perp)$ is the valence-only Fock state of the pion's light-front wave-function.  Diagrams such as those in Fig.\,\ref{instantaneous} connect dynamically-generated chiral-symmetry breaking components of the meson's light-front wave-function to the matrix elements in Eqs.\,(\ref{LFfpi}), (\ref{LFrhopi}).  There are infinitely many contributions of this type and they do not depend sensitively on the current-quark mass in the neighborhood of the chiral limit.  This mechanism is kindred to that discussed in Ref.\,\cite{Casher:1974xd}.

\section{Epilogue}
We have emphasized that absolute confinement of gluons and quarks is a
prerequisite for the containment of condensates within hadrons.  Hence, no
model without confinement, even if treated correctly, can undermine the
foundations of the paradigm that we, and others, are developing for QCD.
The implications of our point are significant and wide-ranging.  For example,
in connection with the cosmological constant, putting QCD condensates back into
hadrons reduces the mismatch between experiment and theory by a factor of
$10^{46}$.  Furthermore, if technicolor-like theories are the correct scheme
for explaining electroweak symmetry breaking, then the impact of the
notion of in-hadron condensates is far greater still \cite{Brodsky:2009zd}.

\begin{acknowledgments}
CDR and PCT acknowledge invaluable collaborations with M.~Bhagwat, L.~Chang, I.\,C.~Clo\"et, A.~Krassnigg, Y.-x.~Liu, P.~Maris, M.\,A.~Pichowsky, S.\,M.~Schmidt and L.~von Smekal, without which significant portions of the progress reported and recapitulated herein would not have been possible.
This work was supported in part by:
U.\,S.\ Department of Energy contract no.~DE-AC02-76SF00515;
U.\,S.\ Department of Energy, Office of Nuclear Physics, contract no.~DE-AC02-06CH11357;
and the U.\,S.\ National Science Foundation, under grants
NSF-PHY-09-69739
and NSF-PHY-0903991.
\end{acknowledgments}



\end{document}